\documentclass[useAMS,usenatbib]{mn2e}

\usepackage{graphicx}	% Including figure files
\usepackage{amsmath}	% Advanced maths commands
\usepackage{amssymb}	% Extra maths symbols
\usepackage{multicol}        % Multi-column entries in tables
\usepackage{bm}		% Bold maths symbols, including upright Greek
\usepackage{pdflscape}	% Landscape pages
\usepackage{multirow,multicol} 

\title[The coolest X-ray emitting gas in galaxy clusters]{Insights into the location and dynamics of the coolest X-ray emitting gas in clusters of galaxies}
\author[C. Pinto et al.]{C. Pinto,$^{1}$\thanks{E-mail:
cpinto@ast.cam.ac.uk} A. C. Fabian,$^{1}$ A. Ogorzalek,$^{2,3}$ I. Zhuravleva,$^{2,3}$ N. Werner,$^{2,3}$  \newauthor
J. Sanders,$^{4}$ Y.-Y. Zhang,$^{5}$ Liyi Gu,$^{6}$ J. de Plaa,$^{6}$ J. Ahoranta,$^{7}$ A. Finoguenov,$^{7}$ \newauthor 
R. Johnstone,$^{1}$ R.E.A. Canning,$^{2,3}$ \\
$^{1}$Institute of Astronomy, Madingley Road, CB3 0HA Cambridge, United Kingdom\\
$^{2}$Kavli Institute for Particle Astrophysics and Cosmology, Stanford University, 452 Lomita Mall, Stanford, CA 94305-4085, USA\\
$^{3}$Department of Physics, Stanford University, 382 via Pueblo Mall, Stanford, CA 94305-4060, USA\\
$^{4}$Max-Planck-Institut f\"ur extraterrestrische  Physik, Giessenbachstrasse 1, 85748 Garching, Germany\\
$^{5}$Argelander-Institut f\"ur Astronomie, Universit\"at Bonn, Auf dem H\"ugel 71, 53121 Bonn, Germany\\
$^{6}$SRON Netherlands Institute for Space Research, Sorbonnelaan 2, 3584 CA Utrecht, The Netherlands\\
$^{7}$Department of Physics, University of Helsinki, 00014 Helsinki, Finland
}
\begin{document}

\date{Accepted 2016 June 13. Received 2016 June 9; in original form 2016 April 29}

\pagerange{\pageref{firstpage}--\pageref{lastpage}} \pubyear{2016}

\maketitle

\label{firstpage}

\begin{abstract}
We extend our previous study of the cool gas responsible for the emission 
of O\,{\small VII} X-ray lines in the cores of clusters and groups of galaxies. 
This is the coolest X-ray emitting phase  
and connects the 10,000\,K H\,$\alpha$ emitting gas to 
the million degree phase, providing a useful tool to understand cooling in these objects.
We study the location of the O\,{\small VII} gas and 
its connection to the intermediate Fe\,{\small XVII} and hotter O\,{\small VIII} phases. 
We use high-resolution X-ray grating spectra of 
elliptical galaxies with strong Fe\,{\small XVII} line emission and 
detect O\,{\small VII} in 11 of 24 objects. 
Comparing the O\,{\small VII} detection level and resonant scattering,
which is sensitive to turbulence and {temperature}, 
suggests that O\,{\small VII} is preferably found 
in cooler objects, where the Fe\,{\small XVII} resonant 
line is suppressed due to resonant scattering, 
indicating {subsonic} turbulence.
{Although a larger sample of sources and further observations is needed to 
distinguish between effects from temperature and turbulence,} 
our results are consistent with cooling being suppressed at high turbulence 
as predicted by models of AGN feedback, gas sloshing and galactic mergers.
In some objects the O\,{\small VII} resonant-to-forbidden line ratio 
is decreased by either resonant scattering or
charge-exchange boosting the forbidden line, as we show for NGC\,4636.
Charge-exchange indicates interaction between neutral and ionized gas phases.
The Perseus cluster also shows a high Fe\,{\small XVII} 
forbidden-to-resonance line ratio, which can be explained with resonant scattering 
by low-turbulence cool gas in the line-of-sight.
\end{abstract}


\section{Introduction}
\label{sec:intro}

The intracluster medium (ICM) embedded in the deep gravitational well
of clusters of galaxies has a complex multi-temperature structure 
with different cospatial phases ranging from $\sim10^{6}$ to above $10^{8}$\,K. 
It is thought to contain most of the baryonic mass of the clusters 
and its density strongly increases in their cores 
where the radiative cooling time is less than 1\,Gyr
and therefore shorter than their age. 
In the absence of heating, this would imply the cooling
of hundreds of solar masses of gas per year below $10^{6}$\,K \citep{Fabian1994}. 
The gas is expected to produce prominent emission lines 
from O\,{\small VI} in UV, peaking at $T\sim3\times10^{5}$\,K, 
as well as O\,{\small VII} ($T\sim2\times10^{6}$\,K) 
and Fe\,{\small XVII} ($T\sim6\times10^{6}$\,K) in X-rays,
suggesting that spectroscopy is a key to understand the cooling processes
in clusters of galaxies.
Evidence of weak O\,{\small VI} UV lines was found by \cite{Bregman2005,Bregman2006}
at levels of $30\,M_{\odot}$\,yr$^{-1}$ or lower,
significantly less than the predicted $100\,M_{\odot}$\,yr$^{-1}$. 
Fe\,{\small XVII} emission lines have been discovered, 
but with luminosities much lower than expected from cooling-flow models 
(see e.g. \citealt{Peterson2003}). 
O\,{\small VII} lines were detected for the first time in a stacked spectrum 
of a sample of cool objects by \cite{Sanders2011} and more recently in individual 
elliptical galaxies by our group \citep{Pinto2014}, but in most cases their fluxes 
are lower than those predicted by cooling flow models.
There is an overal deficit of cool ($\lesssim0.5$\,keV 
or $\lesssim6\times10^{6}$\,K) gas in the cores of clusters 
of galaxies and nearby elliptical galaxies.

Several energetic phenomena are occurring 
in the cores and in the outskirts of clusters of galaxies and isolated galaxies
such as feedback from active galactic nuclei (AGN, see e.g.
\citealt{Churazov2000,McNamara2007,Fabian2012}). Briefly, energetic AGN outflows
drive turbulence in the surrounding ICM, which then dissipates and heats the ICM 
balancing the cooling \citep[see e.g.][]{Zhuravleva2014}.
{AGN can also heat the surrounding gas via dissipation of sound waves
(see e.g.  \citealt{Fabian2003_waves,Fabian2005_waves}).}
The phenomenology can be more complex because galactic mergers and
sloshing of gas within the gravitational potential also produce high turbulence
(see e.g. \citealt{Ascasibar2006}; \citealt{Lau2009}).

In this work we study the coolest X-ray emitting gas in clusters and groups of galaxies
and in elliptical galaxies, which is crucial to understand the ICM cooling from $10^8$\,K 
down to $10^4$\,K. We use high quality archival data and new observations 
taken with the high-resolution Reflection Grating Spectrometer (RGS) aboard XMM-\textit{Newton}.
We search for a relationship between the cool O\,{\small VII} gas and the turbulence,
evidence of resonant scattering and charge exchange in the ICM where neutral gas is observed.
We present the data in Sect.\,\ref{sec:data} and the spectral modeling in  
Sect.\,\ref{sec:spectral_modeling}. We discuss the results in Sect.\,\ref{sec:discussion} 
and give our conclusions in Sect.\,\ref{sec:conclusion}.

\begin{table*}
\caption{XMM-\textit{Newton}/RGS observations used in this paper, extraction regions and O\,{\small VII} detection.}  
% \vspace{-0.2cm}
\label{table:log}      % is used to refer this table in the text
\renewcommand{\arraystretch}{1.1}
 \small\addtolength{\tabcolsep}{-2pt}
 
\scalebox{1}{%
\begin{tabular}{c c c c c c c c c c}     
\hline  
Source                 & t\,$^{(a)}$& $d\,^{(b)}$ &  W\,$^{(b)}$    & CIE\,$^{(c)}$ & $<kT>$\,$^{(c)}$ & $N_{\rm H}$\,$^{(d)}$ & $R_{(f/r) | \rm Fe}$\,$^{(e)}$ & W(O\,{\small VII})\,$^{(e)}$ & P(O\,{\small VII})\,$^{(e)}$ \\  %Slit O\,{\small VII}}$^{max}$
                       & (ks)       &  (Mpc)      &  (')            & (Nr)  & (keV)            &  ($10^{20}\,{\rm cm}^{-2}$) & & (') & ($\sigma$)   \\    %  Ned_avg
\hline                                                  
{{A 262}} (NGC 708)    &   172.6    &   63.7      &    0.20         &  2    & $ 1.19 \pm 0.02$ & 7.15  & $1.91  \pm 0.89 $ &   ---   & ---        \\      %  0109980101/0601 0504780101/0201        0.0161  
Centaurus ({{A 3526}}) &   152.8    &   51.2      &    0.25         &  2    & $ 1.17 \pm 0.02$ & 12.2  & $1.53  \pm 0.22 $ &   0.4   & 3.0        \\      %  0046340101 0406200101                  0.0103  
Fornax ({{NGC 1399}})  &   123.9    &   17.8      &    0.72         &  2    & $ 1.21 \pm 0.03$ & 1.56  & $1.27  \pm 0.18 $ &   ---   & ---        \\     % 0012830101 0400620101                    0.0046  
Perseus  ({{A 426}})   &   162.8    &   72.3      &    0.18         &  2    & $ 2.71 \pm 0.15$ & 20.7  & $4.30  \pm 2.10 $ &   0.8   & 2.7        \\      % 0085110101/0201 0305780101              0.0183  
Virgo (M 87)           &   129.0    &   16.6      &    0.77         &  2    & $ 2.05 \pm 0.05$ & 2.11  & $1.68  \pm 0.19 $ &   ---   & ---        \\      % 0114120101 0200920101                   0.0042  
{{HCG 62}} (NGC 4761)  &   164.6    &   66.1      &    0.24         &  2    & $ 0.85 \pm 0.01$ & 3.76  & $1.57  \pm 0.21 $ &   ---   & ---        \\      % 0112270701 0504780501 0504780601        0.0140  
{{IC 1459}}            &   145.4    &   24.0      &    0.53         &  2    & $ 0.59 \pm 0.04$ & 1.16  & $1.90  \pm 0.65 $ &   3.4   & 3.3        \\      % 0200130101 {0760870101}                 0.006   
{{M 49}} (NGC 4472)    &    81.4    &   15.8      &    0.81         &  2    & $ 0.88 \pm 0.01$ & 1.63  & $1.70  \pm 0.20 $ &   ---   & ---        \\      % 0200130101                              0.0044  
{{M 84}} (NGC 4374)    &    91.5    &   16.7      &    0.77         &  2    & $ 0.83 \pm 0.07$ & 3.38  & $1.86  \pm 0.17 $ &   0.6   & 4.1        \\      % 0673310101                              0.0034  
{{M 86}} (NGC 4406)    &    63.5    &   16.1      &    0.80         &  2    & $ 0.84 \pm 0.05$ & 2.97  & $2.12  \pm 0.29 $ &   3.4   & 5.5        \\      % 0108260201                              -0.0009 
{{M 89}} (NGC 4552)    &    29.1    &   16.0      &    0.80         &  1    & $ 0.62 \pm 0.08$ & 2.96  & $1.62  \pm 0.25 $ &   3.4   & 3.1        \\      % 0141570101                              0.0009  
{{NGC 507}}            &    94.5    &   59.6      &    0.25$^{f}$   &  2    & $ 1.06 \pm 0.02$ & 6.38  & $1.59  \pm 0.67 $ &   ---   & ---        \\     %  {0723800301}                            0.0165  
{{NGC 533}}            &    34.7    &   61.6      &    0.25$^{f}$   &  2    & $ 0.87 \pm 0.03$ & 3.38  & $2.36  \pm 1.04 $ &   ---   & ---        \\      %  0109860101  ID $^{(a)}$                0.0180  
{{NGC 1316}}           &   165.9    &   19.3      &    0.66         &  2    & $ 0.70 \pm 0.02$ & 2.56  & $1.90  \pm 0.25 $ &   0.8   & 6.9        \\     % 0302780101 0502070201                    0.0059  
{{NGC 1332}}           &    63.9    &   22.9      &    0.56         &  2    & $ 0.66 \pm 0.03$ & 2.42  & $3.01  \pm 0.80 $ &   ---   & ---        \\     % 0135980201                               0.0052  
{{NGC 1404}}           &    29.2    &   19.2      &    0.67         &  2    & $ 0.69 \pm 0.01$ & 1.57  & $2.06  \pm 0.25 $ &   0.6   & 2.6        \\     % 0304940101                               0.0065  
{{NGC 3411}}           &    27.1    &   79.1      &    0.25$^{f}$   &  1    & $ 0.93 \pm 0.02$ & 4.55  & $1.24  \pm 0.36 $ &   ---   & ---        \\      % 0146510301                              0.0152  
{{NGC 4261}}           &   134.9    &   29.9      &    0.43         &  1    & $ 0.71 \pm 0.01$ & 1.86  & $1.60  \pm 0.28 $ &   ---   & ---        \\      % 0056340101 0502120101                   0.0073  
{{NGC 4325}}           &    21.5    &    112      &    0.25$^{f}$   &  2    & $ 0.89 \pm 0.02$ & 2.54  & $1.22  \pm 0.32 $ &   ---   & ---        \\     % 0108860101                               0.0259  
{{NGC 4636}}           &   102.5    &   16.0      &    0.80         &  2    & $ 0.72 \pm 0.01$ & 2.07  & $1.95  \pm 0.09 $ &   3.4   & 5.0        \\     % 0111190101/0201/0501/0701                0.0037  
{{NGC 4649}}           &   129.8    &   16.6      &    0.77         &  1    & $ 0.90 \pm 0.01$ & 2.23  & $1.27  \pm 0.20 $ &   ---   & ---        \\      % 0021540201 0502160101                   0.0037  
{{NGC 5044}}           &   127.1    &   35.8      &    0.36         &  2    & $ 0.89 \pm 0.01$ & 6.24  & $1.44  \pm 0.22 $ &   ---   & ---        \\     % 0037950101 0584680101                    0.0090  
{{NGC 5813}}           &   146.8    &   29.2      &    0.44         &  2    & $ 0.68 \pm 0.01$ & 6.24  & $2.61  \pm 0.43 $ &   3.4   & 3.2        \\     % 0302460101 0554680201/0301/0401          0.0064  
{{NGC 5846}}           &   194.9    &   26.9      &    0.48         &  2    & $ 0.74 \pm 0.01$ & 5.12  & $1.67  \pm 0.35 $ &   0.8   & 3.7        \\     % 0021540101/0501 {0723800101/0201}        0.0061  
\hline                
\end{tabular}}

$^{(a)}$ RGS net exposure time. 
$^{(b)}$ Source distance (average value taken from the Ned database: https://ned.ipac.caltech.edu/) and width of the extraction region. 
$^{(c)}$ Number of thermal components and best-fit temperature for a single isothermal model (see Sect.~\ref{sec:gas_turbulence}). \\
$^{(d)}$ Hydrogen column density (see http://www.swift.ac.uk/analysis/nhtot/).
$^{(e)}$ Fe\,{\small XVII} line ratio, width (W) of the region that maximizes the O\,{\small VII} detection and O\,{\small VII} cumulative significance (P)
with ``---'' referring to significance below 99\% (see Sect.~\ref{sec:search_ovii}).\\
$^{(f)}$ For these objects we had to adopt a larger width for the extraction region
to obtain enough statistics in the Fe\,{\small XVII} lines.
\end{table*}
 
\section[]{The data}
\label{sec:data}

The observations used in this paper are listed in Table~\ref{table:log}. 
Most objects were already included in our recent work \citep{Pinto2015},
but here we only focus on those which exhibit cool gas 
producing Fe\,{\small XVII} emission lines.
The original catalog, also known as the CHEERS sample, consists of 44 nearby, 
bright clusters and groups of galaxies and elliptical galaxies
with a $\gtrsim5\sigma$ detection of the O\,{\small VIII} 1s--2p line at 19\,{\AA}
and with a well-represented variety of strong, weak, and non cool-core objects. 
In addition to the CHEERS sources exhibiting Fe\,{\small XVII} emission,
here we include two cool objects: NGC\,1332 and IC\,1459. 
In NGC\,1332, O\,{\small VIII} was detected just below $5\sigma$, 
however its Fe\,{\small XVII} emission lines are much stronger.
IC\,1459 data were enriched by $\sim120$\,ks new data awarded 
during the AO-14. In total we have 24 sources.

The XMM-\textit{Newton} satellite is provided with two main X-ray instruments: RGS and EPIC
(European Photon Imaging Camera).
We have used RGS data for the spectral analysis and EPIC (MOS\,1 detector) data for imaging.
The RGS spectrometers are slitless and the spectral lines are broadened by the source extent.
We correct for spatial broadening through the use of EPIC/MOS\,1 surface brightness profiles.
We repeat the data reduction as previously done in \citet{Pinto2015},
but with newer calibration files and software versions (available by January, 2016).
All the observations have been reduced with the 
XMM-\textit{Newton} Science Analysis System (SAS) v14.0.0.
We correct for contamination from soft-proton flares 
with the standard procedure.

The sources in our sample span a large range of distances (Table~\ref{table:log}).
Therefore, we tried to extract spectra in slices with widths of the same physical size.
Before chosing an absolute scale, we have tested several extraction regions.
For the nearby objects, such as NGC\,4636 
which is the nearest X-ray bright giant elliptical galaxy, 
we adopted a width of about 0.8' ($\sim4$\,kpc)
because it provides a good coverage of the inner Fe\,{\small XVII} bright core,
strengthens the Fe\,{\small XVII} lines with respect to those
produced by the hotter gas phase,
and maximizes the detection of the O\,{\small VII} emission lines.
The spectra of all objects were then extracted in regions centered 
on the Fe\,{\small XVII} emission peak with widths scaled 
by the ratio between the distance of the objects and that of NGC\,4636. 
For NGC 507, 533, 3411 and 4325 we had to adopt a slightly larger width because 
it was the minimum to provide enough statistics in the Fe\,{\small XVII} lines.
The spectra extracted in these regions of approximately equal physical size
have been used to measure the Fe\,{\small XVII} line ratios.
Finally, we have also extracted spectra in different regions, with widths up to 3.4' 
which is the RGS sensitive field of view, to improve the O\,{\small VII} detection.

We subtracted the model background spectrum, which is created by the standard RGS pipeline 
and is a template background file based on the count rate in CCD\,9.
The spectra were converted to SPEX\footnote{www.sron.nl/spex} format through the SPEX task \textit{trafo}.
We produced MOS\,1 images in the $8-27$\,{\AA} wavelength band and extracted
surface brightness profiles to model the RGS line spatial broadening
with the following equation: $\Delta\lambda = 0.138 \, \Delta\theta \, {\mbox{\AA}}$
(see the XMM-\textit{Newton} Users Handbook).

\section{Spectral analysis}
\label{sec:spectral_modeling}

\subsection{Baseline model}
\label{sec:baseline_model}

Our analysis focuses on the $8-27$\,{\AA} first and second order RGS spectra.
We perform the spectral analysis with SPEX
version 3.00.00. We scale elemental abundances to the proto-Solar abundances 
of \citet{Lodders09}, which are the default in SPEX, 
use C-statistics and adopt $1\,\sigma$ errors.

We have described the ICM emission with an isothermal plasma model 
of collisional ionization equilibrium (\textit{cie}). 
The basis for this model is given by the mekal model, 
but several updates have been included (see the SPEX manual).
Free parameters in the fits are the emission measure $Y=n_{\rm e}\,n_{\rm H}\,dV$, 
the temperature $T$, and the abundances (N, O, Ne, Mg, and Fe).
Nickel abundance was coupled to iron.
Most objects required two \textit{cie} components
(see Table\,\ref{table:log}). 
Here, we coupled the abundances of the two \textit{cie} components
and assumed that the gas phases have the same abundances 
because the spectra do not allow us to measure them separately.
The \textit{cie} emission models were corrected for redshift, 
Galactic absorption, see Table~\ref{table:log},
and line-spatial-broadening through 
the multiplicative \textit{lpro} component
that receives as input the MOS\,1 surface brightness profile
(see Sect.\,\ref{sec:data}).

We do not explicitly model the cosmic X-ray background in the RGS spectra 
because any diffuse emission feature would be smeared out into a broad continuum-like component. 
For several objects, including the Perseus and Virgo clusters, 
we have added a further power-law emission component
to account for any emission from the central AGN
{(see \citealt{Russell2013} and references therein)}.
This is not convolved with the spatial profile because 
it is produced by a point source.
For each source, we have simultaneously fitted the spectra of individual observations
by adopting the same model, apart from the emission measures of the \textit{cie} components
which were uncoupled to account for the different roll angles of the observations.

We have successfully applied this multi-temperature model to the RGS spectra.
However, as previously shown in \citet{Pinto2015},
the model underestimates the 17\,{\AA} Fe\,{\small XVII} 
line peaks and overestimates its broadening
for some sources, e.g. Fornax, M\,49, M\,86, NGC\,4636, and NGC\,5813.
This is due to the different spatial distribution of the gas 
responsible for the cool Fe\,{\small XVII} emission lines and 
that producing most of the high-ionization Fe-L and O\,{\small VIII} lines. 
The Fe\,{\small XVII} gas is indeed to be found predominantly in the cores
showing a profile more peaked than that of the hotter gas. 
The spatial profiles estimated with MOS\,1 images strongly depend on the emission 
of the hotter gas due to its higher emission measure
and therefore they overestimate the spatial broadening of the 15--17\,{\AA} lines.
It is difficult to extract a spatial profile for these lines 
because MOS\,1 has a limited spectral resolution
and the images extracted in such a narrow band will lack the necessary statistics 
(see e.g. \citealt{Sanders2013}). 
The 17\,{\AA}\,/\,15\,{\AA} line ratio is also affected 
by resonant scattering (see e.g. \citealt{Gilfanov1987, Sanders2008}), 
which requires a different approach. 
In Sect.\,\ref{sec:gas_location} and \ref{sec:gas_turbulence} 
we account for the different location of the different phases 
and the Fe\,{\small XVII} (and O\,{\small VII}) resonant scattering. 

\subsection{Search for O\,{\small VII}}
\label{sec:search_ovii}

Following \citet{Pinto2014}, we have removed the {O\,{\small VII}} ion from the model
and fitted two delta lines fixed at 21.6\,{\AA} and 22.1\,{\AA}, 
which reproduce the {O\,{\small VII}} resonance and forbidden lines, respectively.
The intercombination line at 21.8\,{\AA} is generally weak or insignificant 
and blends with the resonance line.
These lines are corrected by the redshift, the Galactic absorption,
and the spatial line broadening as done for the \textit{cie} models.
If the resonant line was comparable or stronger than the forbidden lines,
we have determined the {O\,{\small VII}} total significance 
by fixing the resonance-to-forbidden line flux ratio 
to $(r/f) = 1.3$ as predicted by the thermal model.
Otherwise the {O\,{\small VII}} total significance was calculated
as the squared-sum of the significance of each line. 
The latter refers to Perseus, M\,89, and NGC\,4636 and 5813.
We applied this technique to spectra extracted in regions of different widths
in order to search for that one maximizing the {O\,{\small VII}} detection.
We adopt as threshold for the {O\,{\small VII}} detection the 99\% confidence level
because the objects are distributed in two subsamples with detection levels 
$<2.0\sigma$ and $>2.6\sigma$ showing a gap in between.
The results are reported in Table\,\ref{table:log} 
and discussed in Sect.~\ref{sec:discussion}.

\subsection{The location of the cool gas}
\label{sec:gas_location}

It is possible to probe the extent of the cool ({O\,{\small VII}} $-$ Fe\,{\small XVII}) 
gas by comparing its linewidth to that of the hot ({O\,{\small VIII}} $-$ Fe\,{\small XVIII+})
gas. The dominant line broadening effect in grating spectra is indeed produced by the spatial
extent of the source (normally a few $1000$\,km\,s$^{-1}$), which is almost an order of magnitude
larger than the thermal + turbulent broadening (few $100$s\,km\,s$^{-1}$, see e.g. 
\citealt{Pinto2015} and references therein).
The turbulence and thermal broadening are not expected to be 
significantly different between the two phases (see e.g. \citealt{Pinto2015}).
We therefore did the same exercise for the Fe\,{\small XVII} emission lines by
removing the Fe\,{\small XVII} ion from the model and fitting four delta lines fixed at
15.01\,{\AA}, 15.26\,{\AA}, 16.78\,{\AA}, and 17.08\,{\AA},
which are the main Fe\,{\small XVII} transitions.
We do not tabulate the significance of the Fe\,{\small XVII} lines because
they are typically much larger than $5\sigma$.

The \textit{lpro} model in SPEX that corrects for the line broadening 
has an additional scale parameter \textit{s}, which allows to fit the
width of the spatial broadening by a factor free to vary (see the SPEX manual).
We therefore use one \textit{lpro} model to account for the spatial broadening
in the \textit{cie} components that produce the high-temperature lines and 
another \textit{lpro} model to fit the spatial broadening
of the low-temperature {O\,{\small VII}} and Fe\,{\small XVII} lines.
Averaging between all objects in our sample, we find that 
the \textit{lpro} scale parameter of the cool gas
is half of that measured for the hot gas.

\begin{figure}
  \includegraphics[width=0.7\columnwidth, angle=90]{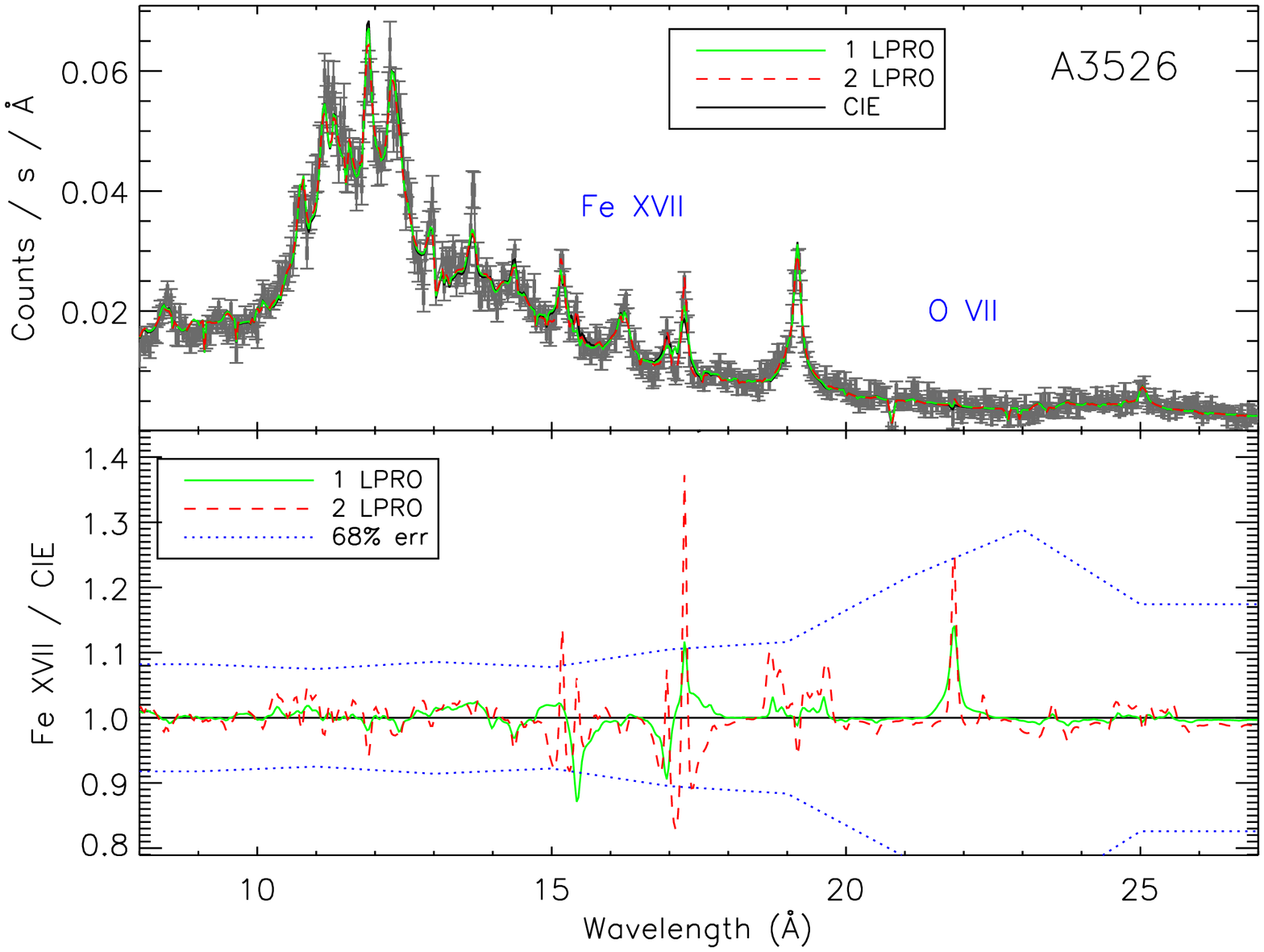}
  \includegraphics[width=0.7\columnwidth, angle=90]{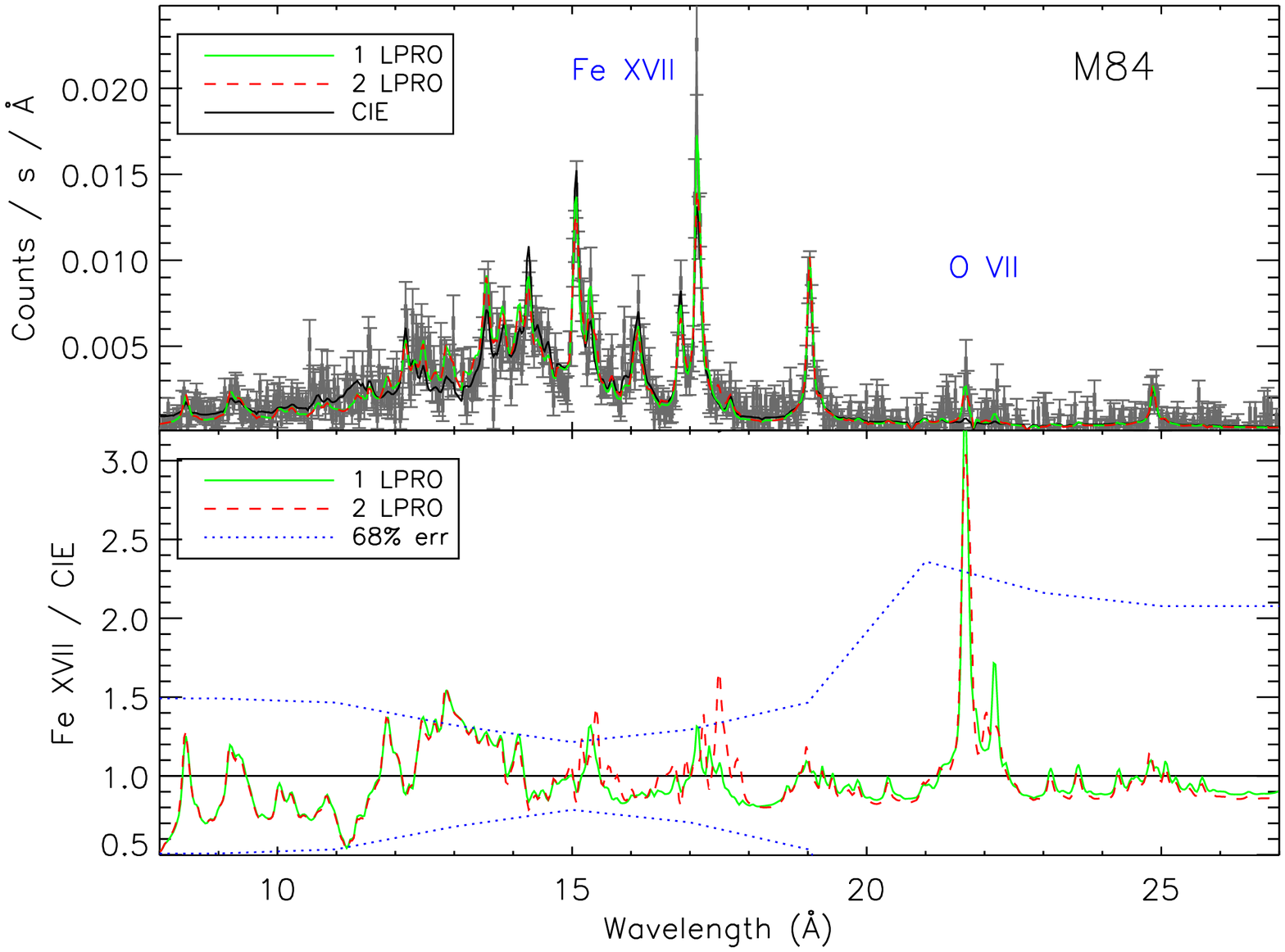}
  \includegraphics[width=0.7\columnwidth, angle=90]{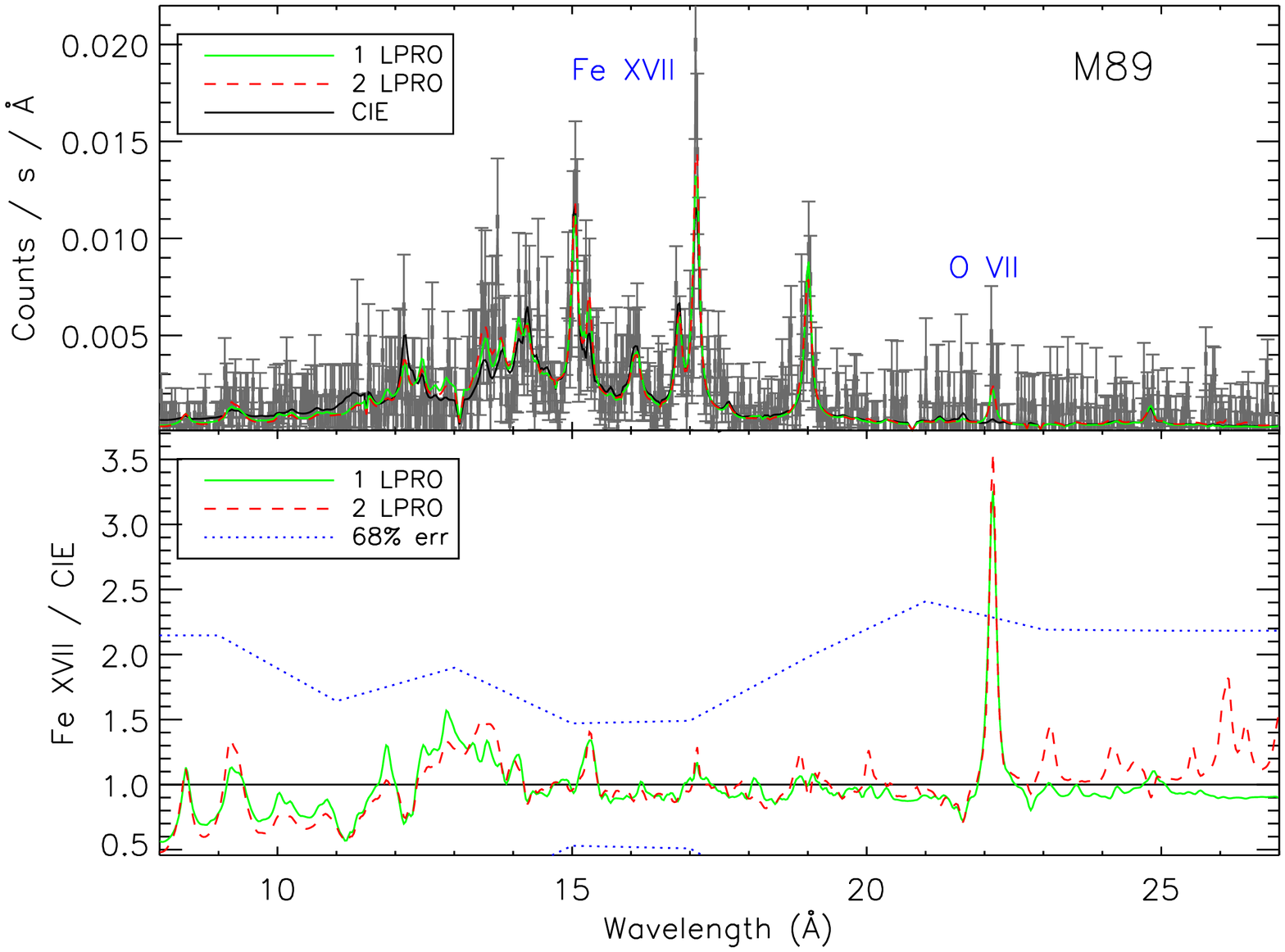}
   \caption{From top to bottom: RGS spectra of the Centaurus cluster,
   M\,84, and M\,89. Three spectral models are overlaid:
   2-\textit{cie} model (thick black line), delta-line Fe\,{\small XVII} model
   (thick green line) and different spatial broadening (dashed red line).
   The bottom panels show the ratios between the Fe\,{\small XVII} line models
   and the 2-\textit{cie} model. The blue dotted lines show the $1\sigma$
   uncertainties.} \label{Fig:Spectra_ratios}
\end{figure}

In Fig.~\ref{Fig:Spectra_ratios} we show the RGS spectra of three interesting sources,
the Centaurus cluster, M\,84 and M\,89 (from top to bottom). 
Overlaid on the data there are three spectral models:
the baseline \textit{cie} model (thick black line), the delta line model 
for {O\,{\small VII}} and Fe\,{\small XVII} lines adopting the same spatial
broadening as the \textit{cie} models (solid green line),
and finally the {O\,{\small VII}} and Fe\,{\small XVII} lines with 
the spatial scale parameter \textit{s} free to vary (dashed red line).
In order to better visualize the effect on the fit from spatial broadening,
we calculate the ratios from the best-fit models obtained with the delta lines
and the best-fitting \textit{cie} model and display them in the bottom panel
of each figure. The color is coded similarly to the top panel: 
the green line is the ratio between the {O\,{\small VII}}--Fe\,{\small XVII} delta
model and the \textit{cie} components (with the same spatial broadening); 
the red line shows the same ratio but with a different spatial broadening.

The Fe\,{\small XVII} lines appear clearly
narrower then the hot--gas lines in the Centaurus cluster (A\,3526),
even taking into account the slightly different thermal broadening,
but there is no significant wavelength shift 
(in agreement with \citealt{Sanders2008}).
This suggests that the Fe\,{\small XVII} cool ($\sim6\times10^6$K) 
gas peaks in the central regions
and has a smaller extent than that of the hot gas responsible 
for the {O\,{\small VIII}} at 19\,{\AA} and the higher-ionization 
Fe\,{\small XX+} ($\gtrsim10^7$K) lines between $11-13$\,{\AA}.
A similar trend is observed in Fornax, M\,49, M\,84, 
NGC\,4636--5044--5813, and Perseus.
The M\,84 and M\,89 elliptical galaxies, whose spectra are dominated 
by the Fe\,{\small XVII} lines, show an {O\,{\small VII}} excess 
with respect to the two-phase \textit{cie} model.
Interestingly, in M\,84 (and NGC\,5846) the {O\,{\small VII}} resonant 
line at 21.6\,{\AA} is in excess, 
while in M\,89 (and NGC\,4636) the excess is shown 
by the forbidden line at 22.1\,{\AA}.
The quality of the spectra of the other objects is not good enough to detected
{O\,{\small VII}} in excess to that already produced 
by the two-\textit{cie} model.
A stronger forbidden line may indicate resonant 
scattering for both {O\,{\small VII}} and Fe\,{\small XVII} 
lines as we previously suggested in \citet{Pinto2014}. 

\begin{figure*}
  \includegraphics[width=1.5\columnwidth, angle=90, bb=65 80 530 680]{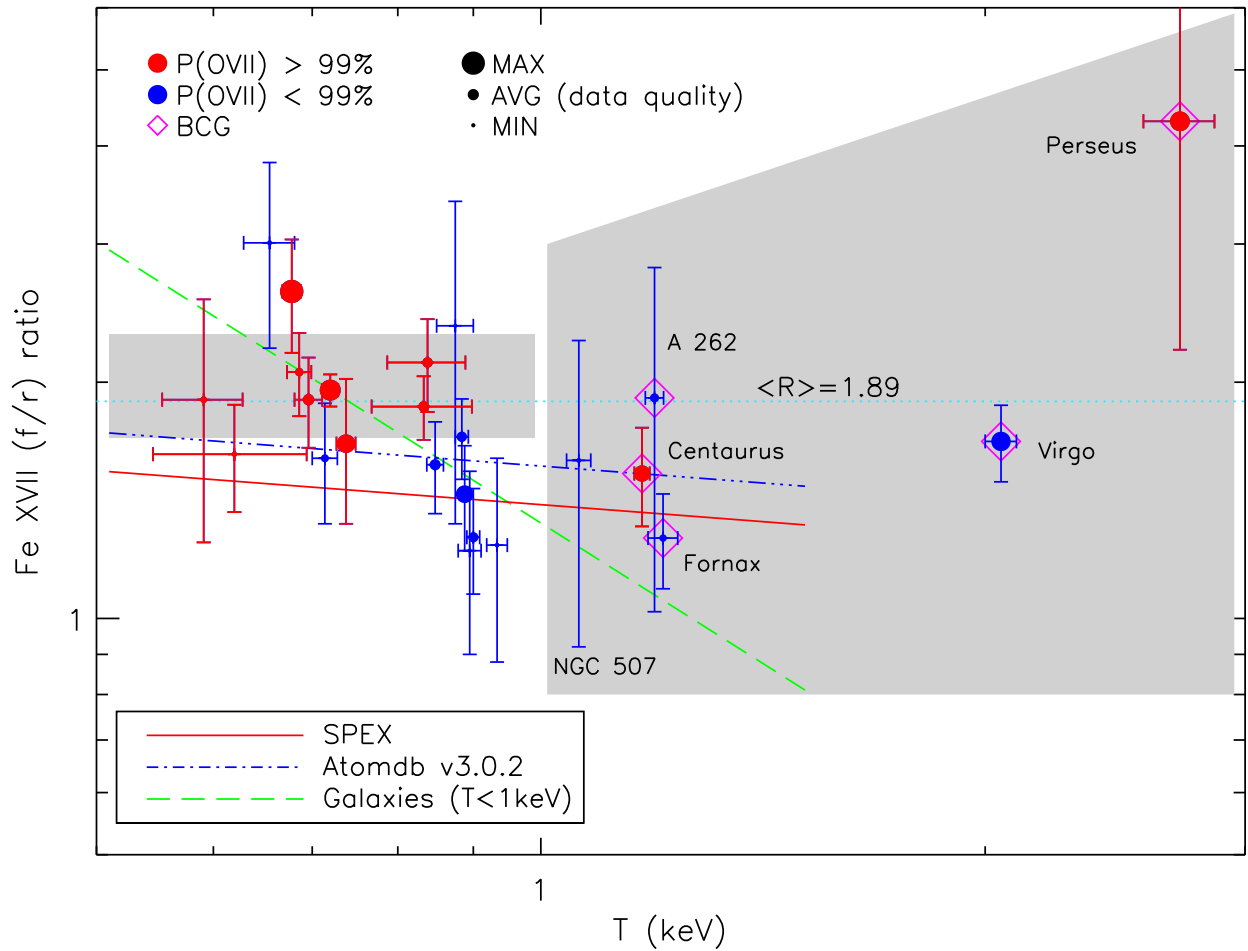}
   \caption{Fe\,{\small XVII} forbidden-to-resonance line ratio versus
   average temperature. O\,{\small VII} detections are reported with red points.
   The dashed green line shows the best fit in the log-log space for the objects below 1\,keV.
   The theoretical predictions from SPEX and Atomdb v3.0.2 are also shown.
   The objects above 1 keV are grey-shaded because there is little Fe\,{\small XVII} at those
   average temperatures and most of it should be produced by a cooler phase.
   The small grey box shows the average Fe\,{\small XVII} ratio ($2.00\pm0.29$)
   of the elliptical galaxies below 1\,keV with O\,{\small VII} detected above 
   the 99\% confidence level.} 
   \label{Fig:Resonant_scattering}
\vspace{-0.4cm}
\end{figure*}

\subsection{O\,{\small VII} VS Turbulence}
\label{sec:gas_turbulence}

When turbulence is low the resonant line can be optically thick;  
it is therefore absorbed and re-emitted in a random direction with the line 
being suppressed towards the bright core and enhanced outside. 
This does not occur at high turbulence
due to the energy shift of the transitions
(see e.g \citealt{Werner2009} and \citealt{dePlaa2012}). 
The forbidden lines have a smaller oscillator strength and are much less affected.
It is then interesting to measure the Fe\,{\small XVII} line resonant scattering 
of the sources in our sample, which is an indicator of 
(low-) turbulence, and compare it to the O\,{\small VII} detection,
in a certain temperature range. 

We have used the Fe\,{\small XVII} line fluxes measured in Sect.~\ref{sec:gas_location}
to calculate the Fe\,{\small XVII} (f/r) line ratios for the models 
with a different spatial broadening between these lines and the hot gas,
and quote the results in Table\,\ref{table:log}.
In order to estimate an average temperature for each source, 
we fit again the RGS spectra with only one single \textit{cie} component.
The average temperatures estimated through these models are quoted in Table\,\ref{table:log}.

We plot the Fe\,{\small XVII} (f/r) line ratios versus the temperature 
in Fig.~\ref{Fig:Resonant_scattering} with the red points showing the sources 
with O\,{\small VII} detection above the 99\% confidence level.
The point size scales with the average S/N ratio of the RGS spectra at 17\,{\AA}.
We also show the Fe\,{\small XVII} line ratios as predicted by a thermal
model without resonant scattering according to the Atomdb v3.0.2 and SPEX
to visualize the strength of the resonant scattering in each source
and the systematic uncertainties in the atomic data.
Points below the theoretical predictions would be unphysical because
it is difficult to strengthen only the resonant line in the line-of-sight
towards the center of the galaxies (although charge-exchange can slightly enhance 
the Fe\,{\small XVII} resonant line). All our Fe\,{\small XVII} (f/r) line ratios
agree with the theoretical predictions or are above the theoretical curves,
indicating moderate resonant scattering and therefore low-to-mild (subsonic) turbulence.
We fit a straight-line in the log-log space for the objects with $T<1$\,keV, 
and found a significant anti-correlation between the Fe\,{\small XVII} (f/r) line ratio 
and the average temperature (well above the $3\sigma$ confidence level, 
$p-$value$=0.00026$, with slope $-0.79\pm0.18$, 
see the green dashed line in Fig.~\ref{Fig:Resonant_scattering}).
This may indicate either a decrease in optical depth or 
an increase in turbulence or both.
We caution against the comparison with the brightest cluster galaxies (BCG), 
i.e. those in A\,262, Centaurus, Fornax, Perseus, and Virgo clusters, 
because above 1\,keV the ICM becomes optically thin to the 
Fe\,{\small XVII} emitted by the cooler gas phases ($kT<0.9$\,keV)
and therefore resonance scattering becomes insensitive to turbulence 
in the cores of these systems.
All results are discussed in Sect.~\ref{sec:discussion}.

\subsection{Systematic effects}
\label{sec:systematics}

{There are several systematics that may affect our results and their interpretation 
such as the background subtraction, the line blending and the uncertainties in the atomic database.}

{The model background spectra used throughout this work adopt long exposures of blank fields.
This is a safe approach since any background contribution to the weak O\,{\small VII} 
or the strong Fe\,{\small XVII} lines would be smeared out in a continuum like feature.
For some bright and compact objects such as NGC\,1316--1404 M\,89
we could extract a background spectrum in the outer regions of the RGS detector
and match it with the model background spectrum.
The spectra were comparable and no significant difference in the line ratios were found.
We also tested a different continuum with a local (14.5--18.0\,{\AA}) fit 
using a power-law and a few delta lines obtaining larger statistical uncertainties 
and consistent results with the previous Fe\,{\small XVII} measurements.
We tested a power-law continuum for the 19.5--22.5\,{\AA} range 
obtaining similar O\,{\small VII} detections.}

{We have also checked the effects of blending with other lines. 
The O\,{\small VII} resonant and forbidden lines 
are located in a rather clean spectral range apart from O\,{\small VI} and O\,{\small VII} 
intercombination lines. As mentioned in Sect.\,\ref{sec:intro}, there are only small amounts 
of O\,{\small VI} in these objects as clearly shown in far-UV spectra. 
The O\,{\small VI} stronger line at 22.0\,{\AA} is also expected to be resolved by RGS
due to the smaller extent, and therefore line broadening, of the O\,{\small VI-VII} cool phases. 
The O\,{\small VII} intercombination line is 5.5 weaker than the resonant line 
and also not expected to significantly affect our results. 
The Fe\,{\small XVII} resonant and forbidden lines are in a crowded spectral region, 
but they are much stronger than the neighbor lines. 
We have artificially doubled the flux of the brightest neighbor lines, which is more
than the statistical uncertainties. The Fe\,{\small XVII} (f/r) line ratio was consistent 
with the standard measurements.}

{The uncertainties in the atomic database do not affect our measurements of line ratios,
but of course the interpretation of resonant scattering. There is a significant ($>20$\%)
difference between the Fe\,{\small XVII} (f/r) line ratio as predicted by AtomDB and SPEX.
This means that we do not know the absolute value of resonant scattering in our sources,
which is crucial to estimate the absolute scale of turbulence, but the relative differences 
between line ratios measured in different objects should not be highly affected.}

\section{Discussion}
\label{sec:discussion}

In Sect.\,\ref{sec:search_ovii} we have searched for O\,{\small VII} ($\sim 2 \times 10^6$\,K) 
gas in a sample of 24 objects, including clusters and groups of galaxies and elliptical galaxies,
with strong ($>5\sigma$) Fe\,{\small XVII} line emission.
We have detected O\,{\small VII} above the 99\% confidence level in 11 sources
{and shown that O\,{\small VII} is preferably found in the cores of the sources,
possibly following the distribution of the Fe\,{\small XVII} ($> 5 \times 10^6$\,K) gas.
Exceptions are IC\,1459 and M\,89 where the lower count rate requires to integrate
photons over a larger region. For M\,86, NGC\,4636 and NGC\,5813
the O\,{\small VII} is also better detected in the wider slit most likely due to 
their more extended cool cores.

In order to search for a link between cooling and turbulence, we have plotted the
Fe\,{\small XVII} forbidden-to-resonant line ratio with the temperature and the
O\,{\small VII} significant detections in Fig.\,\ref{Fig:Resonant_scattering}.
The high quality data points show some evidence 
for O\,{\small VII} to be mainly detected in sources 
with significant resonant scattering,
which indicates the low level of turbulence.
Although our sample is incomplete and the resonant scattering 
is more sensitive at lower temperatures,
our results are consistent with a picture where turbulence is heating the gas 
and preventing it to cool below $\sim0.45$\,keV, where O\,{\small VII} 
line emission begins to be important.

\subsection{O\,{\small VII} charge exchange or scattering?} \label{sec:ngc4636}

At temperatures of 0.2-to-0.6\,keV the O\,{\small VII} 
resonance-to-forbidden line ratio is predicted to be between 1.25--1.35. 
We found (r/f) line ratios lower than 1.25 in the RGS spectra of 
NGC\,4636, M\,89, and NGC\,1404 as already shown in \citet{Pinto2014}.
Our values could be due to either suppression of the resonant line 
via resonant scattering or enhancement of the forbidden line
by photoionization or charge exchange.

The O\,{\small VII} resonance line at 21.6\,{\AA} may be subject 
to resonant scattering. At the temperature of $\sim0.5$\,keV, 
where the Fe\,{\small XVII} ionic concentration peaks, the O\,{\small VII} 
is optically thin and no longer self-absorbed along the line-of-sight.
However, it is possible that the gas is distributed in various
non-volume filling phases at different temperatures.
We multiplied the two \textit{cie} emission components for a collisionally-ionized absorbing model
(\textit{hot} model in SPEX) and re-fit the RGS spectrum of NGC\,4636.
We obtained a column density of $1.24\pm0.30\times10^{20}\,{\rm cm}^{-2}$ 
with a temperature of $0.23\pm0.03$\,keV, which is lower than the 
$0.43\pm0.07$\,keV value measured for the \textit{cie} component responsible
for the O\,{\small VII} emission.
The presence of such cool gas is suggested by the detection of a large amount 
of H$\alpha$ emission in the core of NGC\,4636 \citep{Werner2014}.
It is suspicious, however, that the absorbing gas is cooler than the emitting gas 
despite the need to be located (on average) in outer regions where
higher temperatures are expected unless the cool gas is clumpy.

The astrophysical processes that strengthen the 
O\,{\small VII} forbidden line emission
are photoionization and charge exchange. We can rule out photoionization
because no bright AGN is observed in NGC\,4636. 
Charge exchange (CX hereafter) occurs when ions interact 
with neutral atoms or molecules; 
one or more electrons are transferred to the ion into an excited state,
which decays and emits a cascade of photons 
increasing the forbidden-to-resonance ratios of triplet transitions.
This process is often observed in supernova remnants 
(e.g. Puppis A, \citealt{Katsuda2012}), 
starburst galaxies (e.g. M\,82, \citealt{Liu2011}) 
and colliding stellar winds (e.g. Solar Wind, \citealt{Snowden2004}).

The CX plasma code recently provided by \citet{Gu2016} 
is implemented in SPEX v3.00.00 (\textit{cx} model). \citet{Gu2015} first used this code 
to successfully describe the unidentified 3.5\,keV feature 
in the lower resolution CCD spectrum of the Perseus cluster.
We re-fit the NGC\,4636 spectrum 
with a new \textit{cie} (driven by the Fe\,{\small XVII-XVIII} lines) 
+ \textit{cx} (mainly, O\,{\small VII-VIII}, Ne\,{\small X}, and Mg\,{\small XI})
model corrected by redshift and Galactic absorption and obtain results 
comparable to the resonant scattering (\textit{hot}) model described above.
In the fit we exclude the $13.8-15.5$\,{\AA} spectral range because it contains
several Fe\,{\small XVII} lines suppressed by resonant scattering
which would lead to a wrong estimate of the temperature.
In Fig.\,\ref{Fig:NGC4636} we show the best fit with the contribution 
from the \textit{cie} and \textit{cx} components. Charge exchange provides a reasonable
description of the O\,{\small VII} lines and produces significant
O\,{\small VIII}, Ne\,{\small X}, and Mg\,{\small XI} emission and
accounts for $\sim10\%$ of the flux in the 0.3--2.0\,keV energy band.

\begin{figure}
  \includegraphics[width=0.7\columnwidth, angle=-90,bb=50 100 510 752]{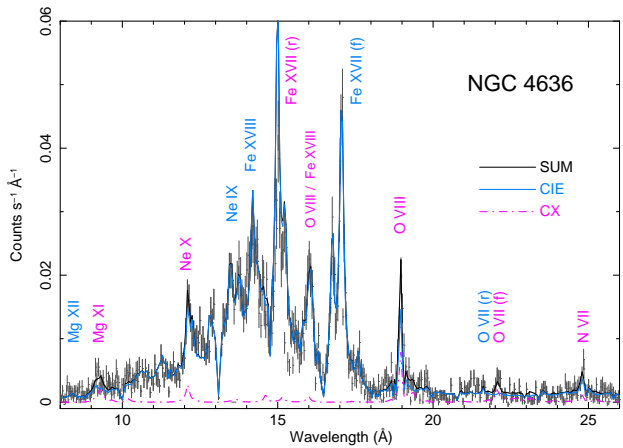}
   \caption{NGC\,4636 RGS first order spectrum with hybrid model
   consisting of isothermal and charge-exchange components.} \label{Fig:NGC4636}
\end{figure} 

The ionic temperature of the \textit{cx} component was coupled to the $\sim0.7$\,keV 
temperature of the \textit{cie} component.
If left free to vary, a better fit provides $T_{\rm ion}=0.40\pm0.05$\,keV,
in agreement with the 2-\textit{cie} model, which may suggest that the charge exchange is 
occurring between neutrals and the cooler O\,{\small VII} phase rather than 
the hotter gas phase associated with the Fe\,{\small XVII} lines. 
This may indicate that the cool O\,{\small VII} gas is a better
tracer of the cold neutral phase and that they could be somewhat cospatial,
both distributed in clumps. 
The CX code calculates velocity-dependent rates with which
we measure a collision velocity lower than 50\,km\,s$^{-1}$ (at 68\% level),
in agreement with the low turbulence found in NGC\,4636 \citep{Werner2009}.
This is the first time that a charge exchange model is successfully applied
to a high-resolution X-ray spectrum of a giant elliptical galaxy.

\subsection{Resonant scattering in Perseus?}

In Fig.\,\ref{Fig:Resonant_scattering} we have shown that the Perseus cluster has an
unexpected, high ($4\pm2$), Fe\,{\small XVII} (f/r) line ratio. 
The spectrum extracted within a larger region of width $\sim0.8'$ 
(see Fig.\,\ref{Fig:Perseus}) holds much smaller error bars 
and constrains Fe\,{\small XVII} (f/r) $\geq4$.
This value is higher than that measured in any other object and 
remarkable if compared to the other clusters 
(A\,262, Centaurus, Fornax and Virgo). 
The inner core of the Perseus cluster is dominated by a hot $\sim3$\,keV plasma, but
it has been clearly shown to be multiphase with the inner arcminute ($\sim20$\,kpc) 
having significant emission from 0.5--4\,keV \citep[see e.g.][]{Sanders2007}.

Below 1\,keV and in a low-turbulence regime the 15\,{\AA} resonance line is optically
thick and it may therefore be subject to resonant scattering in the line-of-sight.
We have therefore re-fitted the Perseus spectrum multiplying the two thermal components
by a collisionally-ionized absorption model (\textit{hot} model) to test the
suppression of the Fe\,{\small XVII} resonant line 
(as previously done for the O\,{\small VII} lines 
in NGC\,4636 in Sect.\,\ref{sec:ngc4636}). 
We have ignored the first order spectra between 10 and 14\,{\AA} due to high pile up 
and use the second order RGS 1 and 2 spectra because they are not 
significantly affected by pileup and their statistics peak in this wavelength range.
This model reasonably describes the 15--17\,{\AA} Fe\,{\small XVII} lines 
(see Fig.\,\ref{Fig:Perseus}) and provides a column density of 
$\sim2\times10^{20}\,{\rm cm}^{-2}$ and a temperature of $\sim0.6$\,keV.

\citet{Fabian2015} suggested that high-resolution X-ray spectra enable to search
for evidence of ICM absorption onto the AGN continuum in NGC\,1275, 
the brightest cluster
galaxy in Perseus, with a focus on the hard X-ray band where Fe\,K lines dominate.
We have tested the same approach in the soft RGS band by applying the \textit{hot}
absorption model only to the nucleus, which was fitted with a power law;
the two \textit{cie} emission line components are only absorbed by the Galactic neutral ISM.
This AGN-only absorption model is statistically indistinguishable to the previous one,
with $\Delta\chi^2$ and $\Delta$C-stat of 6 for 1948 degrees of freedom,
but a column density of $\sim1.5\times10^{21}\,{\rm cm}^{-2}$ is required, 
in good agreement with the predictions of \citet{Fabian2015}. 
{If the suppression of the 15\,{\AA} Fe\,{\small XVII} resonant line 
and the detection of absorption are interpreted as resonant scattering, 
which is a very likely scenario, then this means that the cool gas 
in Perseus is characterized by low turbulence.}

\begin{figure}
  \includegraphics[width=0.7\columnwidth, angle=-90,bb=80 63 540 675]{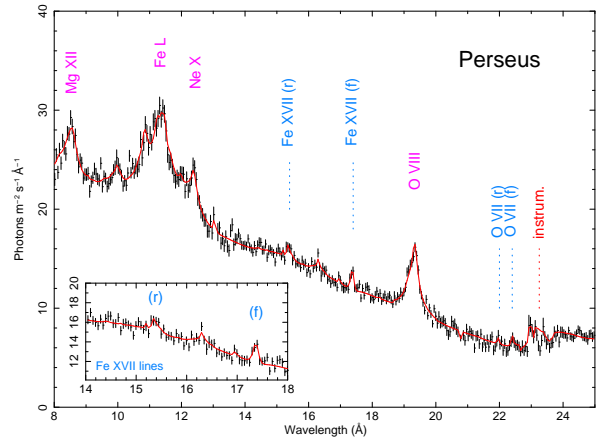}
   \caption{Perseus RGS first order spectrum with multiphase thermal emission model
   absorbed by isothermal gas at 0.6\,keV.} \label{Fig:Perseus}
\end{figure}

\section{Conclusions}
\label{sec:conclusion}

In this work we have confirmed and extended our previous 
discovery of O\,{\small VII} emission lines 
in spectra of elliptical galaxies as well as groups and clusters of galaxies.
This is the coolest X-ray emitting intracluster gas 
and seems to be connected to the mild Fe\,{\small XVII} gas,
being located preferably at small (1-10\,kpc) scales.
The O\,{\small VII} is often detected in objects with strong resonant
scattering of photons in the Fe\,{\small XVII} lines, 
indicating {subsonic} turbulence.
This would be consistent with a scenario where cooling is suppressed by turbulence 
in agreement with models of AGN feedback, gas sloshing and galactic mergers.
{We note that a larger sample of sources and consequently more observations
are needed to better disentangle resonant scattering effects 
due to temperature and turbulence; the current sample is incomplete.}
In some objects the O\,{\small VII} resonant line is weaker than the forbidden line
either due to resonant scattering or to charge-exchange processes 
occurring in the gas as we have shown for NGC\,4636.
The Perseus cluster shows an anomalous, high, Fe\,{\small XVII} 
forbidden-to-resonance line ratio, which can be explained with resonant scattering 
by cool gas in the line-of-sight under a regime of low turbulence.
In two forthcoming papers (Ogorzalek et al., Pinto et al.) 
we will compare the measurements of Fe\,{\small XVII} line ratios with those
predicted by theoretical models of resonant scattering that take into account thermodynamic 
properties of these objects in order to estimate the turbulence in both their cores
and outskirts. This will provide further insights onto the link between cooling, turbulence,
and the phenomena of AGN feedback, sloshing, and mergers occurring in
clusters and groups of galaxies.

\section*{Acknowledgments}

This work is based on observations obtained with XMM-\textit{Newton}, an
ESA science mission funded by ESA Member States and USA (NASA).
We also acknowledge support from ERC Advanced Grant Feedback 340442
and new data from the awarded XMM-\textit{Newton} proposal ID 0760870101.
Y.Y.Z. acknowledges support by the German BMWi through the
Verbundforschung under grant 50OR1506.

\bibliographystyle{aa}
\bibliography{bibliografia} %----> bibliografia.bib

\bsp

\label{lastpage}

\end{document}